\documentclass[twocolumn,trackchanges]{aastex631}

\usepackage{lineno}
\usepackage{booktabs}

\begin{document}

\title{Beyond the Local Void: A data-driven search for the origins of the Amaterasu particle}

\correspondingauthor{Nadine Bourriche}
\email{nadineb@mpp.mpg.de}
\email{capel@mpp.mpg.de}

\author[0009-0004-1555-8004]{Nadine Bourriche}
\affiliation{Max Planck Institute for Physics \\
Boltzmannstra\ss e 8, 85748 Garching, Germany}
\affiliation{Technical University of Munich \\ 
James-Franck-Stra\ss e 1, 85748 Garching}

\author[0000-0002-1153-2139]{Francesca Capel}
\affiliation{Max Planck Institute for Physics \\
Boltzmannstra\ss e 8, 85748 Garching, Germany}

\begin{abstract}

We introduce a simulation-based inference framework to constrain the origins of individual ultra-high-energy cosmic rays by combining realistic three-dimensional propagation modeling with Bayesian parameter estimation. Our method integrates \texttt{CRPropa 3} simulations, including all relevant interactions and magnetic deflections in both Galactic and extra-Galactic fields, with Approximate Bayesian Computation to infer posterior distributions over key parameters such as source position, distance, energy, and magnetic field properties. This approach allows joint constraints from the observed energy and arrival direction to be applied simultaneously, naturally incorporating their correlations in addition to relevant modelling uncertainties.
We demonstrate our method by applying it to the Amaterasu particle detected by the Telescope Array observatory, the second-highest-energy cosmic ray ever detected. The resulting posterior distributions quantify the regions of space consistent with its reconstructed properties under different energy and composition assumptions, revealing a broader set of nearby source candidates than found in previous analyses. This application highlights the framework’s ability to translate individual UHECR observations into directly interpretable source constraints and provides a foundation for future simulation-based analyses of cosmic rays at the highest energies.

\end{abstract}


\section{Introduction}\label{sec:intro}

Ultra-high energy cosmic rays (UHECRs) are charged particles with energies that exceed $E \geq 10^{18}$\,eV. The origin of these particles is still unknown despite numerous efforts to identify their sources. These searches are challenged by the complex nature of UHECR propagation, including energy losses and deflections by intervening magnetic fields. Although advancements in the search for sources of UHECRs have been made, including the possible association of UHECRs with starburst galaxies (SBGs) and active galactic nuclei (AGN) catalogs, these models struggle to explain UHECR observations at the highest energies, beyond $10^{20}$\,eV \citep{PAO2018,TA2018,Capel2019,Halim2024}. While few in number, events at these energies have the potential to be the most constraining in terms of their origins. Their high energies suggest nearby horizons and trajectories are less affected by magnetic fields, even when considering the evidence for the increasing mass and charge of UHECRs as a function of energy \citep{Batista2019}. As such, studying individual events at the highest energies is a complementary and powerful way to search for the sources of UHECRs \citep{Globus2023,Bourriche2023,Bianciotto:2025pg}.

We present a data-driven approach to map out the volume of the Universe consistent with the origins of individual UHECRs at the highest energies. Our method makes use of Approximate Bayesian Computation (ABC, \citealt{Beaumont2019}) to enable the use of 3D simulations with \texttt{CRPropa 3} \citep{Batista2022} and the inclusion of a comprehensive description of UHECR propagation in our inference. We also include the statistical and systematic uncertainties in the reconstructed properties of UHECR events, as well as key modelling uncertainties regarding the Galactic and extra-Galactic magnetic fields and possible particle energy at the source. The result of applying ABC in this context is a 3D posterior distribution of possible source locations that are consistent with the observed energy and direction of individual UHECR events, marginalising over the various sources of uncertainty. Contrary to previous work, the constraints from both the energy and direction measurement are applied simultaneously, allowing us to include important correlated effects. For example, we expect larger source distances to lead to larger possible particle deflections and longer paths for particles interactions, affecting both the energy and arrival direction of detected events.

We demonstrate our approach through application to the Amaterasu particle observed by the Telescope Array Collaboration (TA, \citealt{TA2023}). Amaterasu was detected by the surface detector of the Telescope Array with an energy of ${E=244\pm29\,\mathrm{(stat.)}^{+51}_{-76}\,\mathrm{(syst.)}}$\,EeV and an arrival direction of ${\mathrm{R.A.},\,\mathrm{Dec.}=(255.9\pm0.6,\,16.1\pm0.5)^{\circ}}$, making it the second-highest energy particle ever detected and the highest energy particle ever detected in a currently operating observatory. The high estimated energy of Amaterasu makes it a strong candidate for an individual event-based UHECR source search. Previous works have investigated the possible source of Amaterasu by studying its compatibility with models for UHECR production in nearby galaxies \citep{Kuznetsov2023} and by estimating its deflection and horizon through backtracking and 1D simulations \citep{TA2023,Unger2024}. The results suggest that Amaterasu's detected direction does not correspond to any known active galaxy but seems to come from the Local Void, an especially low-density region of the Universe \citep{Tully2008}. This conclusion has led to the proposal of past astrophysical transient sources \citep{Farrar2024}, ultraheavy cosmic rays \citep{zhang2024}, magnetic monopoles \citep{Frampton2024}, Lorentz invariance violation \citep{Lang:2024ld,Das:2025gd}, and superheavy dark matter \citep{sarmah2024,Murase:2025pf} as possible explanations for the Amaterasu observation. 

In this work, we use our approach to explore the possibility that Amaterasu's origins extend beyond the Local Void within a standard picture of cosmic ray propagation. The paper is organised as follows: in Section~\ref{sec:simset}, we describe the \texttt{CRPropa~3} simulation setup, the free parameters considered, and our choice of prior assumptions, in Section~\ref{sec:methods} we detail our implementation of ABC to exploit the results of the simulations, in Section \ref{sec:results} we present our results and in Section~\ref{sec:sources} we discuss their implications in terms of possible astrophysical sources.

\section{Physical model} \label{sec:simset}

To model the propagation of UHECRs, we use \texttt{CRPropa~3} to perform 3D simulations. We include all relevant particle interactions: photo-pion production, photo-disintegration, electron-pair production, as well as nuclear decay and adiabatic losses. 

\subsection{Observed energy}

As discussed in \citet{TA2023}, the reconstructed energy reported for the Amaterasu particle is subject to large systematic uncertainties. To take these uncertainties into account in our analysis, we consider two different cases for the detected energy, $E_\mathrm{det}$: 1) the nominal case with ${E_\mathrm{nom} = 244}$\,EeV, and 2) the lower end of the systematic range with ${E_\mathrm{low} = 168}$\,EeV. In both cases, the statistical uncertainties are treated equivalently with $\sigma_E = 29$\,EeV, and we do not consider the upper end of the reported systematic range as these results will only be more constraining than those in case 1). One major source of systematic uncertainty in the TA energy measurements is that in the mass of the particle, estimates for the same event being 10\% lower under the assumption of an iron nucleus than of a proton. The nominal value of 244 EeV is obtained under the assumption of a proton or light nucleus, whereas the lower systematic uncertainty also takes into account the possibility of a heavier nucleus. Nevertheless, we consider a full range of possible arrival masses for the particle in both scenarios 1) and 2).

\subsection{Magnetic fields}

The extra-Galactic magnetic field (EGMF) remains relatively poorly understood. We expect magnetic fields of up to $\sim 1$\,$\mu$G within Galaxy clusters while in filaments they are expected to be below $\sim 10$\,nG, and even lower in voids \citep{Durrer2013,COLEMAN2023102819}. In this work, the EGMF is modelled as a Gaussian random field with a Kolmogorov turbulence spectrum on a regular 3D grid with a spacing of 25\,kpc. The EGMF is parameterised by the root mean square field strength, $B_\mathrm{rms}$, and the coherence length, $L_\mathrm{c}$. Given the expectations for the EGMF of the local environment of the Milky Way and nearby Galaxies considered here (e.g.~\citealt{10.1093/mnras/stx3354,Locatelli:2021se}), we consider ${\left\langle B_\mathrm{rms}^2 L_\mathrm{c} \right\rangle^{1/2} \lesssim 10^{-8} \, \text{G} \, \text{Mpc}^{1/2}}$, where ${L_\mathrm{c} \lesssim 1}$\,Mpc \citep{Kotera2011}. To reflect the large uncertainties while satisfying this constraint, we choose log-uniform priors over the EGMF parameters such that ${B_\mathrm{rms}\sim \log{\mathrm{U}}[0.1,\,10]}$\,nG and ${L_\mathrm{c} \sim \log{\mathrm{U}}[60,\,1000]}$\,kpc. The lower bounds of these priors are chosen such that the EGMF deflections in this range are expected to have a negligible impact on the results. The upper bounds allow for a relatively strong EGMF, but we note that the choice of a log-uniform prior results in a density that is concentrated towards the lower bounds, as shown below in Fig.~\ref{fig:rig_B}.  

The Galactic magnetic field (GMF) is much stronger than the EGMF (on the order of several $\mu$G), and in addition to a turbulent component it includes a large-scale regular component which results in the coherent lensing of UHECR directions. Unfortunately, direct three-dimensional measurements are not possible and inferring its structure from line-of-sight integrated observable quantities requires making hypotheses about its shape, for example based on the structure of the magnetic field observed in external galaxies. A suite of eight models spanning different assumptions on the coherent GMF are presented in \citet{unger2024coherent}, and independent modelling efforts including the impact of the Local Bubble have also been recently reported \citep{Korochkin:2025ne}. These different GMF models present a discrete systematic for the UHECR propagation modelling that is challenging to incorporate in our present framework other than by repeating the analysis for all $\sim 10$ available models. Here, we use the best-fit UF23 base model \citep{unger2024coherent} for the coherent field and the striated and small-scale random turbulent components from the JF12 Planck-tune~\citep{2016A&A...596A.103P} to demonstrate the implementation of our method. We plan to include a range of GMF models in our future work. 

\subsection{Particle properties at the source}

We assume that Amaterasu leaves its source as an iron nucleus. The assumption of iron at the source is well-motivated by the fact that heavier particles are more easily accelerated to higher energies and further enables comparison with previous similar studies \citep{Unger2024,Kuznetsov2023}. While we treat Amaterasu as an individual extreme event, the observed UHECR spectrum also supports models assuming heavy source compositions (see, e.g., \citealt{Halim2024,Unger2024} and references therein). This assumption is further motivated by our goal to explore possible origins beyond the Local Void within a standard UHECR propagation framework, requiring large deflections and therefore a highly charged particle at the source \citep{TA2023}.  In principle, our framework can treat the particle type at the source as a free parameter, but we choose to fix it here due to the lack of constraints available from the current observations and for the clarity of the resulting interpretation.

The energy of Amaterasu at the source is left as a free parameter, $E_\mathrm{src}$. For this parameter, we choose a power-law spectrum prior of $E^{-1}$. This value is the best-fit spectral index obtained by the combined composition and spectrum fit of the Pierre Auger data \citep{Aab2017}. The minimum energy at the source is set to be 3$\sigma_E$ below the reconstructed energy of Amaterasu, $E_\mathrm{det}$. This choice of minimum energy is driven by the fact that arrival energies more than $3\sigma_E$ lower than the detected energy will be rejected by our ABC algorithm. The maximum energy is chosen to be 3 ZeV to explore the impact of trans-GZK \citep{Griesen1966} energies on our results, and we also report a subset of results with ${E_\mathrm{src} \leq 500}$\,EeV for comparison in Appendix~\ref{sec:lower_Emax}.

\subsection{Source location}
\label{sec:source_loc}

To map out the possible volume of space consistent with the measured energy and arrival direction of Amaterasu, we consider the source location as a set of free parameters: the galactic longitude and latitude, $(l,\,b)$, and the distance, $D_\mathrm{src}$. The prior for $D_\mathrm{src}$ is set such that source positions are uniformly sampled within a spherical volume. We also only consider distances in the range $[1,\,D_\mathrm{max}]$ Mpc. The efficiency of our ABC algorithm decreases as we consider larger source distances, meaning that our choice of the maximum source distance in our prior is affected by computational as well as physical considerations. For case 1), we set the corresponding ${D_\mathrm{max} = 12}$~Mpc, and for case 2) we use ${D_\mathrm{max} = 15}$~Mpc. Ideally, we would consider a larger ${D_\mathrm{max}}$, but as discussed further in Sections~\ref{sec:results} and \ref{sec:sources} (see also Fig.~\ref{fig:selection_efficiency}), this will only lead to more possible source associations and further possible origins outside of the Local Void.

Similarly, the ideal choice for the priors on $l_\mathrm{src}$ and $b_\mathrm{src}$ would correspond to a uniform distribution on the sphere, covering the full sky. Unfortunately, due to computational constraints, this choice is not possible with the current implementation of our ABC algorithm. As such, we set up a source direction prior that is as uninformative as possible, while also including relevant information on expected deflections in the GMF and EGMF, as described in the following. 

Firstly, we estimate the impact of the GMF by backtracking \citep{Thielheim2002}. We model the galaxy as a sphere of $20$\,kpc, placing the Earth $8.5$\,kpc from the center. Starting from the detected energy and arrival direction of Amaterasu, 10,000 realizations of an iron nucleus are backtracked to the Galactic boundary, sampling over the statistical uncertainties in the detected particle properties and assuming they are Gaussian. We record the average deflection, $\theta_\mathrm{GMF}$, and the mean direction, $\mu_\mathrm{GMF}$ resulting from the backtracking. 

Secondly, we calculate the expected EGMF deflections, $\theta_\mathrm{EGMF}$, based on the other free parameters of our model, $D_\mathrm{src}$, $B_\mathrm{rms}$ and $L_\mathrm{c}$, according to the following approximation \citep{Harari:2002sl}
\begin{equation}
    \theta \approx 2.3^{\circ} Z \left(\frac{E}{50\,\mathrm{EeV}}\right)^{-1}\left(\frac{B_\mathrm{rms}}{1\,\mathrm{nG}}\right)\left(\frac{D_\mathrm{src}}{10\,\mathrm{Mpc}}\right)^{1/2}\left(\frac{L_c}{1\,\mathrm{Mpc}}\right)^{1/2},
    \label{eqn:emgf_deflections}
\end{equation}
where $Z=26$, as we assume Amaterasu is iron at the source, and we use the detected energy of Amaterasu to give an estimate of maximal EGMF deflections. 

\begin{figure*}[ht]
\centering
\includegraphics[width=0.9\textwidth]{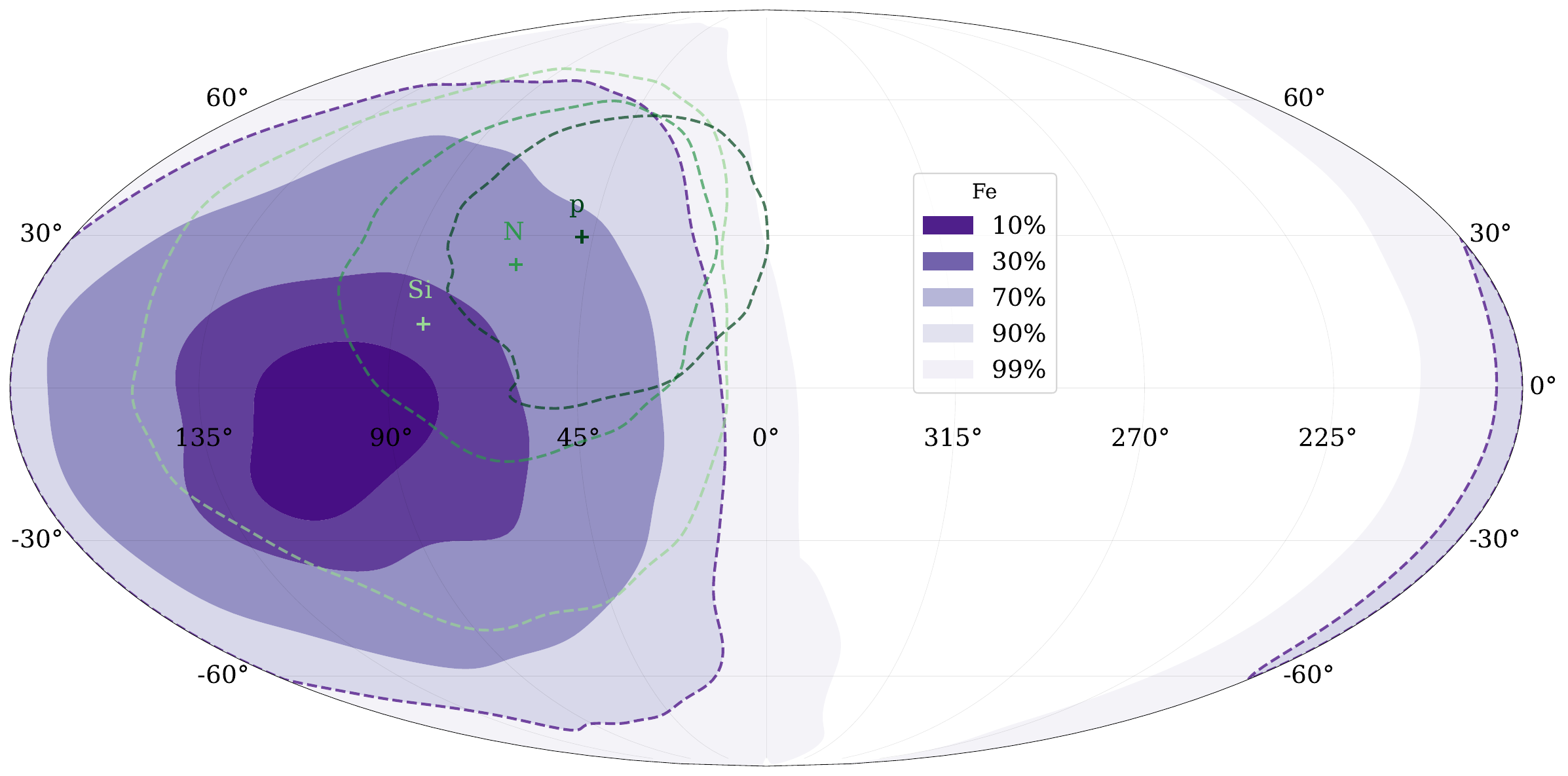}
\includegraphics[width=0.9\textwidth]{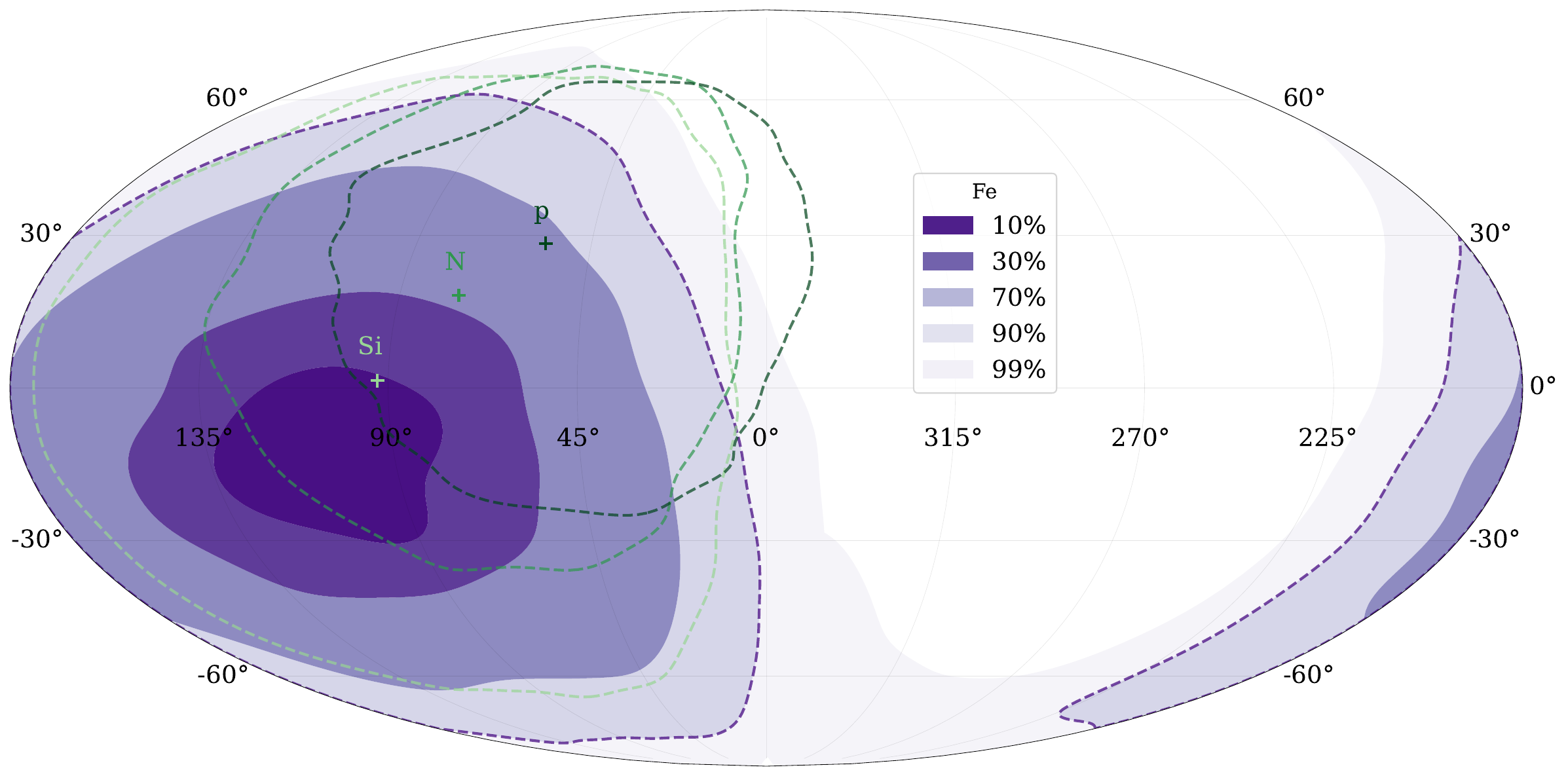}
\caption{Sky maps of the prior on the source position for $E_{\text{nom}}$ (upper panel) and $E_{\text{low}}$ (lower panel). The shaded contours show the regions of highest prior density, for 10, 30, 70, 90 and 99\%. As detailed in the text, our prior assumes that Amaterasu arrives as an iron nucleus when considering the Galactic deflections. For comparison, we also show the mean values and 90\% contours that result from calculating our source direction prior assuming lighter nuclei for the Galactic backtracking. In particular protons, nitrogen and silicon are shown by the dark, mid and light green dashed lines, respectively.}
\label{fig:prior_skymap}
\end{figure*}

Our source direction prior is then defined as a von Mises-Fisher distribution\footnote{The von Mises-Fisher distribution is the equivalent of a Gaussian on the surface of a sphere. It has two parameters: a mean direction, $\mu$, and a shape parameter, $\kappa$, which describes how tightly the distribution is clustered around $\mu$. A uniform distribution corresponds to $\kappa = 0$ and as $\kappa$ increases, the distribution becomes more concentrated around $\mu$.} centred on $\mu_\mathrm{GMF}$ and with a width corresponding to the largest deflection when comparing $\theta_\mathrm{GMF}$ and $\theta_\mathrm{EGMF}$. We also consider a maximum width of the distribution to be $87^{\circ}$, otherwise rejecting the trial. This choice is motivated as containing roughly 70\% of the total probability for a von Mises-Fisher distribution with shape parameter $\kappa = 1$, which is almost isotropic. This allows us to make our computation more efficient while still allowing for large deflections. As the EGMF deflections depend on the other free parameters, the source position prior is slightly different at every iteration of the ABC algorithm. In Fig.~\ref{fig:prior_skymap}, we show the sky maps of the prior distribution for $E_{\text{nom}}$ and $E_{\text{low}}$ resulting from running our ABC algorithm and including this cut. This figure also shows the impact of assuming lighter nuclei for the Galactic backtracking step described above. While the resulting prior distribution covers over half of the sky, we acknowledge that the shape of our prior, in particular the choice of the mean direction from backtracking an iron nucleus, will have some impact on the results shown here. However, we also note that the 99\% probability contour of the used prior (assuming iron) contains the 90\% contours of all lighter nuclei priors for the $E_{\text{nom}}$ case and there is also good coverage of these assumptions in the $E_{\text{low}}$ case. We aim to further quantify the impact of this assumption in future work by improving the computational efficiency of our method. 

\subsection{Summary}

Our physical model has a total of six free parameters: $l_\mathrm{src}$, $b_\mathrm{src}$, $D_\mathrm{src}$, $E_\mathrm{src}$, $B_\mathrm{rms}$ and $L_c$. The prior assumptions for these free parameters are summarised in Table~\ref{tab:priors}.

\begin{table}
\begin{tabular}{llll}
\toprule
\textbf{Param.} & \textbf{Prior dist.} & \textbf{Range} & \textbf{Unit} \\
\midrule
$B_\mathrm{rms}$ & $\log{\mathrm{U}}$ & $[0.1,\,10]$ & nG \\
$L_c$ & $\log{\mathrm{U}}$ & $[60,\,1000]$ & kpc \\
$E_\mathrm{src}$ & $\propto E_\mathrm{src}^{-1}$ & $[E_\mathrm{det}-3\sigma_\mathrm{E},\,3000]$ & EeV \\ 
$D_\mathrm{src}$ & $\propto D_\mathrm{src}^3$ & $[1,\,D_\mathrm{max}]$ & Mpc \\
$(l_\mathrm{src}, b_\mathrm{src})$ & \multicolumn{2}{c}{See Fig.~\ref{fig:prior_skymap}} & deg \\
\bottomrule
\end{tabular}
\caption{Free parameters and their corresponding prior assumptions. $\log{\mathrm{U}}$ denotes a log-uniform distribution. The joint prior for $(l,b)$ is a function of $B_\mathrm{rms}$, $L_c$ and $D_\mathrm{src}$. Further details are given in the text. \label{tab:priors}}
\end{table}

\section{Methods} \label{sec:methods}

We take a simulation-based approach to inference, using ABC to estimate the posterior distribution of our free parameters. The advantage of using ABC in this context, as opposed to traditional methods, is that we do not need to define the likelihood function that connects the observed data to the model parameters explicitly; instead, our choice of simulation defines it implicitly, as defined in Section~\ref{sec:simset}. ABC proceeds by sampling parameter values from the prior distributions, running a simulation based on these parameters to generate a simulated data set, and then comparing this simulated data to the observed data. Proposed parameter values are then accepted or rejected based on how closely the simulated data match the observations, thereby approximating the posterior distribution.

We use the reported arrival direction and reconstructed energy of Amaterasu as data. We define a distance metric for comparison with a simulated event as $\mathrm{|X_{sim}-X_{obs}|}$, where $\mathrm{X_{sim}}$ is the Galactic latitude, longitude or the energy of the simulated event and $\mathrm{X_{obs}}$ is that of the observed one. To compare the two, a tolerance $\mathrm{\epsilon}=3\sigma$ is introduced, with $\sigma$ denoting the reconstruction uncertainty on the event measurements. A set of parameters is accepted if the resulting simulation produces an event with both its arrival energy and direction within $\mathrm{3\sigma}$ of those observed for Amaterasu. Each accepted parameter set is weighted based on how well the resulting events match the observed event. The weights for each observable are defined as $\mathrm{\omega_X = \frac{1}{\sigma_X \sqrt{2\pi}} \exp \left( -0.5 \left( \frac{X_{obs} - X_{sim}}{\sigma_X} \right)^2 \right)}$, and then combined them into a total weight $\mathrm{\omega=\omega_{l}\cdot\omega_{b}\cdot\omega_E}$. 

For each proposed combination of parameters, we simulate as described in Section~\ref{sec:simset}, with $\mathrm{10^6}$ particles emitted isotropically from a single source location. This number gives a good compromise concerning computational efficiency and numbers of detected events. We also consider different realisations of the turbulent fields for both the GMF and EGMF at each iteration of our ABC algorithm to account for this uncertainty into our results. For all the proposed parameter sets resulting from our prior, only a fraction of $\sim 4 \times 10^{-4}$ result in simulations with accepted events, making the implementation of our ABC algorithm computationally challenging. There are rare cases where more than a single event is accepted from a single simulation iteration, forming a fraction of $\sim 10^{-5}$ of all simulation iterations, and $\sim 0.03$ of simulations iterations resulting in at least one accepted event. In these cases, one is randomly chosen to define the resulting weights. 

\section{Results} \label{sec:results}

We apply our ABC approach to the nominal and low energy cases, with the number of resulting parameter sets shown in Table~\ref{tab:acc_param_sets}. To explore the impact of different arrival particle types, the accepted parameter sets are split into three groups based on the mass number of the corresponding accepted events, $A$: light ($A \leq 4$), medium ($4 < A \leq 28$), and heavy ($A > 28$). We summarise the resulting posterior distributions for $l_\mathrm{src}$, $b_\mathrm{src}$, $D_\mathrm{src}$, $E_\mathrm{src}$, $B_\mathrm{rms}$ and $L_\mathrm{c}$ below. In the visualisation of our results, we use Gaussian kernel density estimation, making sure to account for hard parameter boundaries, where relevant. We also verify that the resulting distributions are robust to reasonable variations in the choice of tuning parameters. 

\begin{table}
\begin{tabular}{lllll}
\toprule
& \textbf{Total} & \textbf{Light} & \textbf{Medium} & \textbf{Heavy} \\
\midrule
$E_\mathrm{nom}$ & 272 & 5.9\,\% & 16.2\,\% & 77.9\,\% \\
\textbf{$E_\mathrm{low}$} & 255 & 14.9\,\% & 12.2\,\% & 72.9\,\% \\
\bottomrule
\end{tabular}
\caption{Number of accepted parameter sets resulting from running ABC. The total is also broken down to show the contribution of three different arrival particle categories: light ($A \leq 4$), medium ($4 < A \leq 28$) and heavy ($A > 28$). \label{tab:acc_param_sets}}
\end{table}

\begin{figure*}[ht]
\centering
\includegraphics[width=0.9\textwidth]{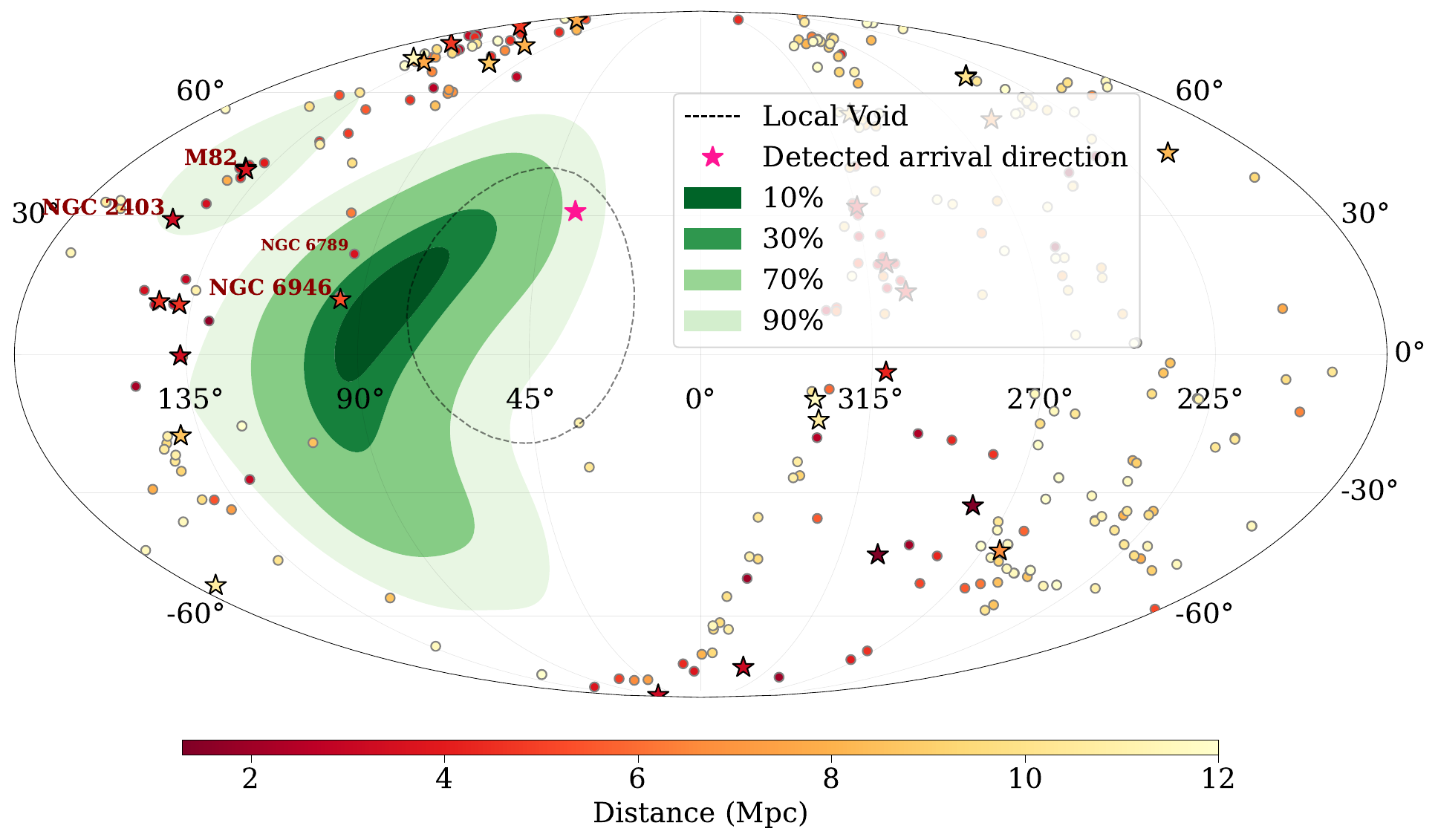}
\includegraphics[width=0.9\textwidth]{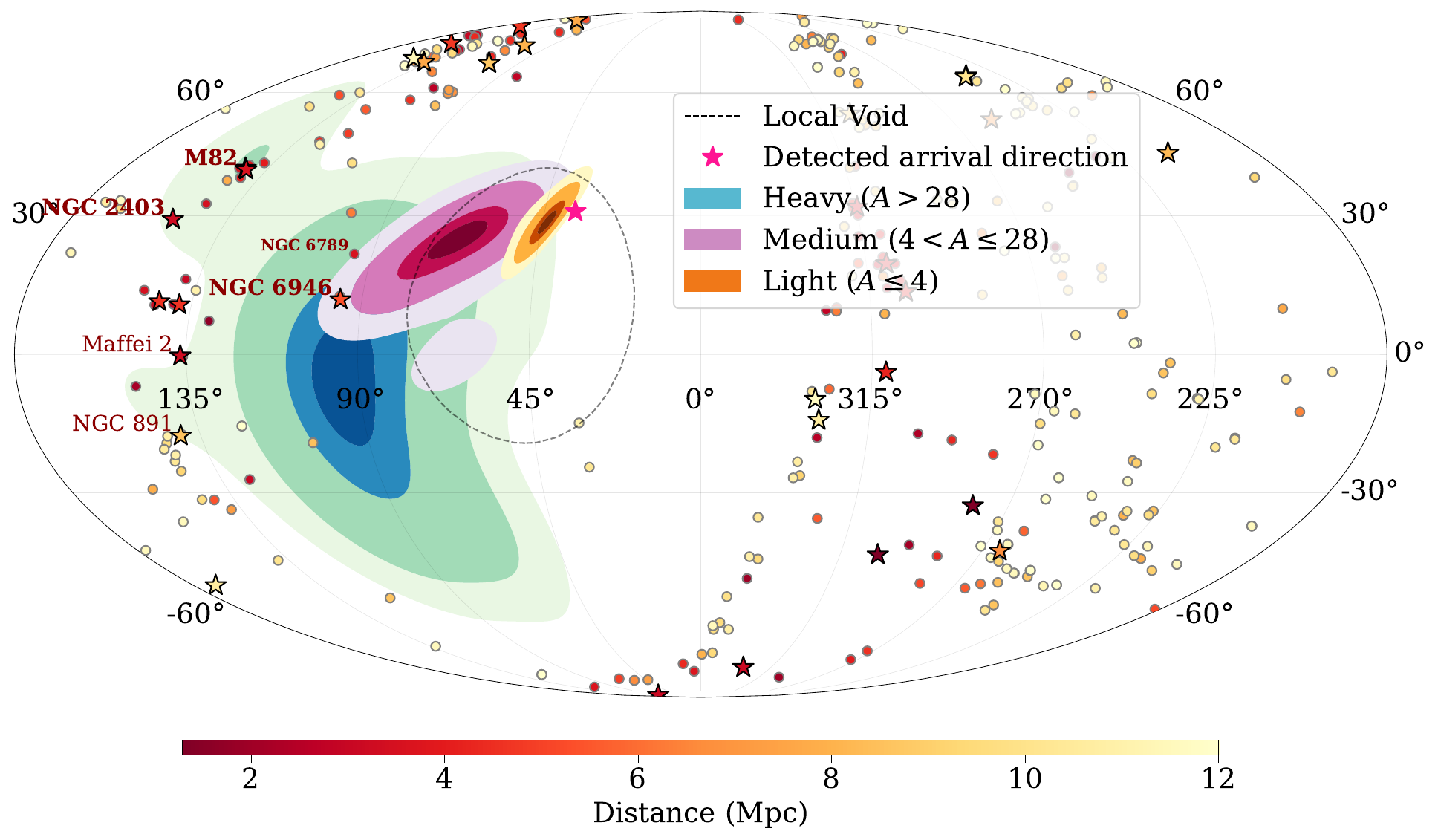}
\caption{Sky maps resulting from the nominal energy run (case 1) showing the possible source positions of Amaterasu in Galactic coordinates. The measured arrival direction of Amaterasu and the outline of the Local Void are shown for reference. The markers show galaxies within the accepted $D_\mathrm{src}$ range, with stars indicating SBGs and AGN, and circles showing quiescent galaxies. In the upper panel, all accepted parameter sets are considered and the green contours outline the labelled regions of highest posterior density. The lower panel shows the arrival mass-dependent posterior distribution, conditional on the particle arriving with a mass number, $A$, in one of three groups. The contour levels are as in the upper panel. \label{fig:Enom_skymap}}
\end{figure*}

Fig.~\ref{fig:Enom_skymap} shows the sky maps in Galactic coordinates for case 1) with ${E_\mathrm{nom}=244\pm29}$~EeV. The total posterior distribution is given in the upper panel, and compared to known astrophysical source candidates. We consider the SBG source list from \citet{PAO2018}, as well as AGN from \citet{Baumgartner2013} and \citet{Ajello2017}. For completenes, we also include quiescent galaxies from the 2MASS survey \citep{Huchra2012}. Only objects within $D_\mathrm{max} = 12$\,Mpc are shown and all points are colour-coded according to their distance from Earth. The lower panel of Fig.~\ref{fig:Enom_skymap} shows the same plot but for the posterior distribution assuming different arrival particle types. As shown in Table~\ref{tab:acc_param_sets}, we find that most of the accepted simulations result in events that are at least as heavy as silicon, with light and medium compositions less likely. This ratio is likely influenced to some extent by our choice for the source direction prior (see Section~\ref{sec:source_loc}). However, we are still able to see results pulling away from this prior choice for the light and medium particle mass cases. Overall, we clearly see that selecting larger arrival masses leads to posterior distributions that are more deflected from the measured Amaterasu arrival direction and also with broader tails, as expected from the combined effect of the GMF and EGMF. 

\begin{figure*}[ht]
\centering
\includegraphics[width=0.9\textwidth]{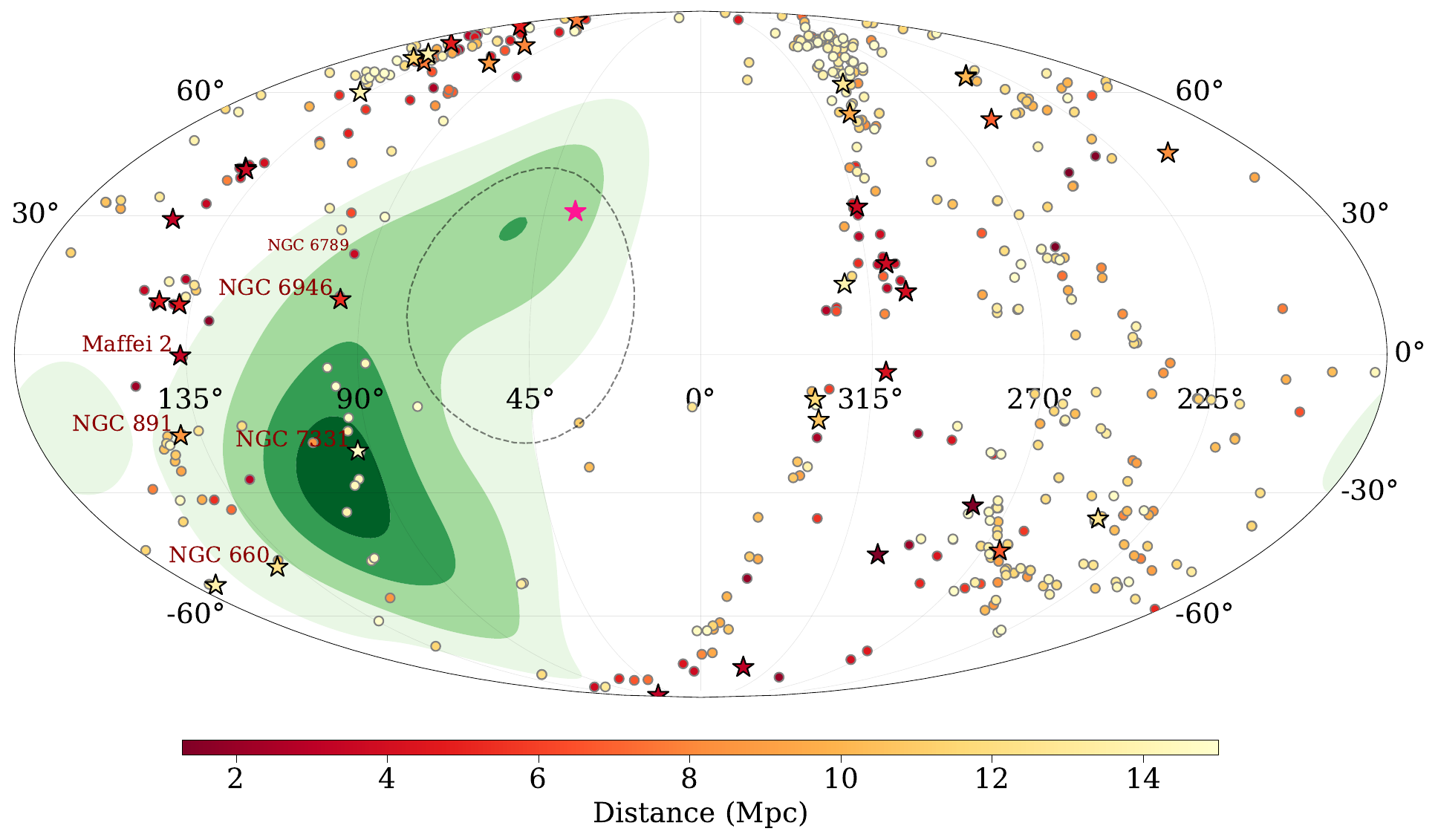}
\includegraphics[width=0.9\textwidth]{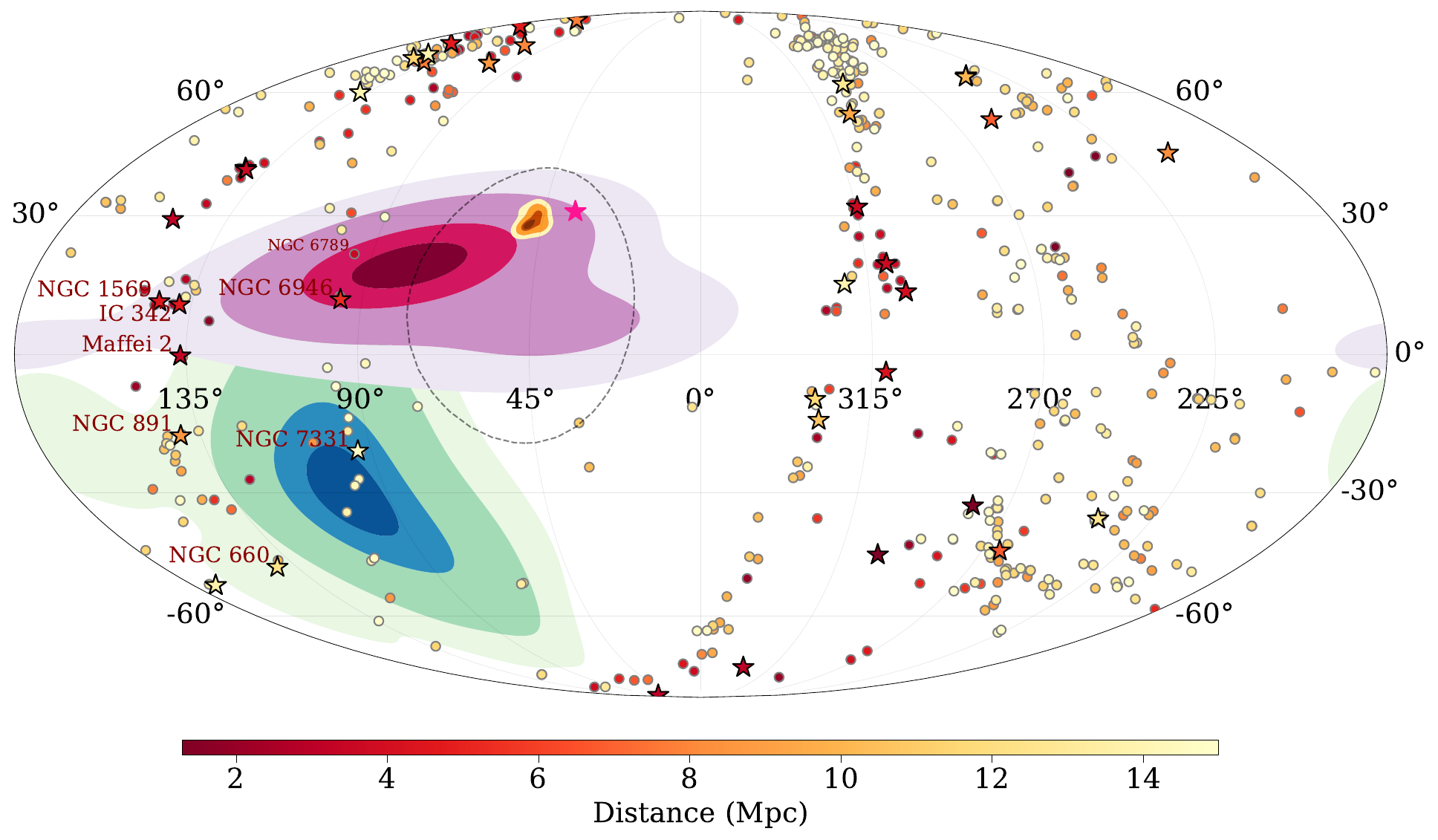}
\caption{Sky maps of the low energy run (case 2). The layout is as in Fig.~\ref{fig:Enom_skymap}, with the upper panel showing the total posterior distribution and the lower panel showing the composition-dependent distributions. \label{fig:Elow_skymap}}
\end{figure*}

The corresponding sky maps for case 2) with ${E_\mathrm{low} = 168 \pm 29}$~EeV are shown in Fig.~\ref{fig:Elow_skymap}. The resulting posterior distributions for the source direction follow a similar trend to those shown in Fig.~\ref{fig:Enom_skymap}, but are more deflected from the measured arrival direction of Amaterasu and further extended in nature since the lower energies considered allow for a stronger impact of both the GMF and EGMF.

\begin{figure}[ht]
\includegraphics[width=\columnwidth]{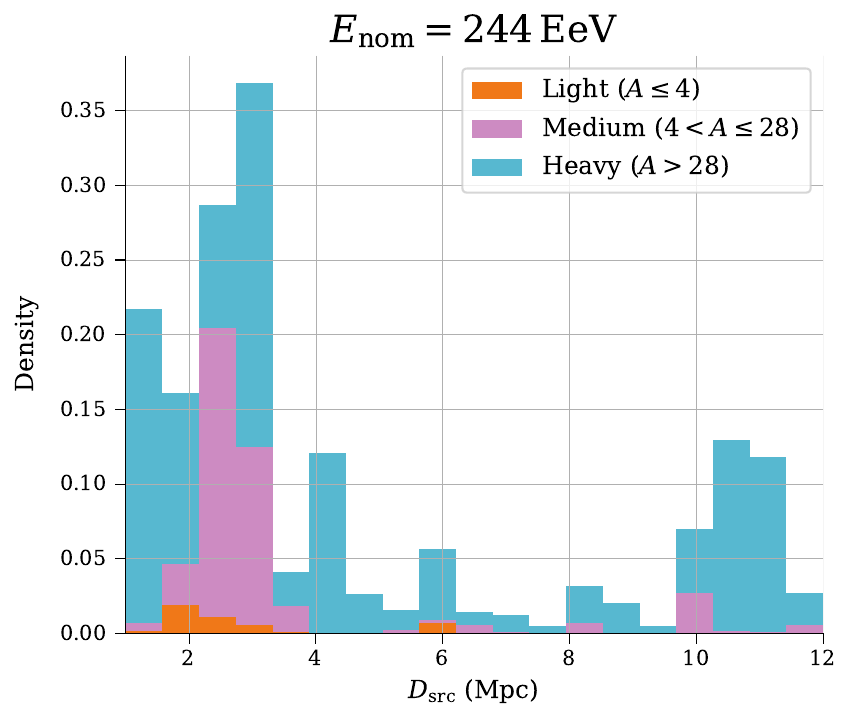}
\includegraphics[width=\columnwidth]{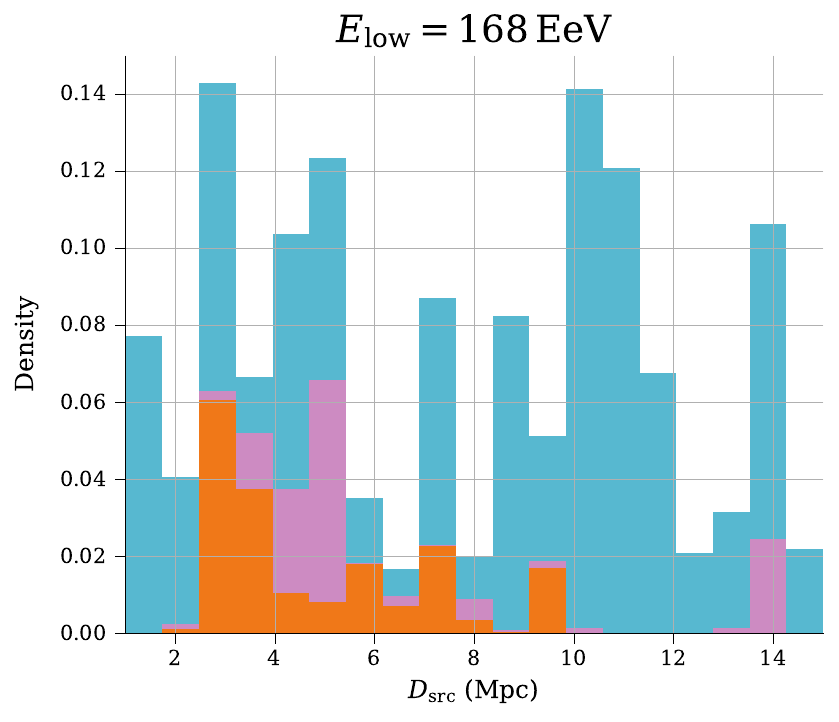}
\caption{The marginal posterior distribution of $D_\mathrm{src}$ is shown for the nominal and low energy cases in the upper and lower panels, respectively. The results are given as a stacked histogram to highlight the relative contributions of $D_\mathrm{src}$ values that lead to accepted events in different arrival mass groups. Orange, pink, and blue sections indicate the fraction of each bar that has arrival particle mass numbers of ${A \leq 4}$, ${4 < A \leq 28}$, and $A > 28$, respectively. The height of the bar gives the total density summed over all mass numbers and the histogram is normalised. \label{fig:Dsrc_comp_hist}}
\end{figure}

Fig.~\ref{fig:Dsrc_comp_hist} shows the marginal posterior distributions of $D_\mathrm{src}$ for the cases of both $E_\mathrm{nom}$ and $E_\mathrm{low}$. The distributions are shown as stacked histograms assuming different arrival mass groups highlighted in different colours, as in the lower panels of Figs.~\ref{fig:Enom_skymap} and \ref{fig:Elow_skymap}. For $E_\mathrm{nom}$, the source distance is relatively constrained, with the most probable distance around ${D_\mathrm{src} \sim 3}$\,Mpc. While the composition is more diverse at closer distances with the presence of light, medium and heavy nuclei, a heavier composition is predominant for ${D_\mathrm{src} > 6}$~Mpc, with another lower peak a distances between 10 and 12\,Mpc. The case of $E_\mathrm{low}$ is more complex, with a most probable distance of ${D_\mathrm{src} \sim 4}$~Mpc for light and medium mass groups, but a longer tail and relatively even density across the distance range for heavier nuclei. 

\begin{figure}
    \centering
    \includegraphics[width=\columnwidth]{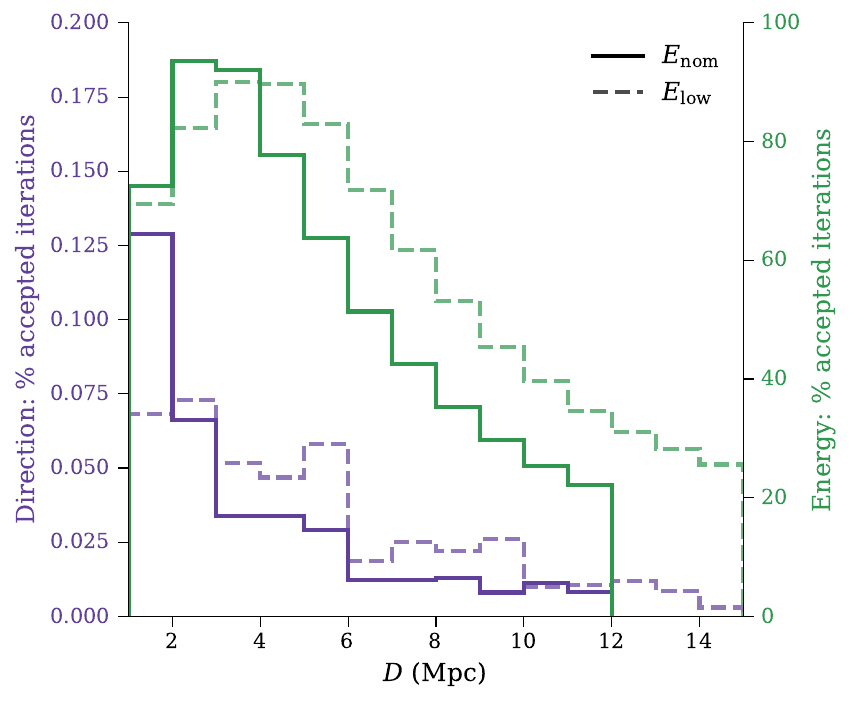}
    \caption{The fractions of accepted iterations of our ABC method are shown for the selections on the arrival direction and the arrival energy. Note the different scales on the left and right y-axes.}
    \label{fig:selection_efficiency}
\end{figure}

The marginal posterior distributions of $D_\mathrm{src}$ shown here are a result of a joint selection on the energy and arrival direction of simulated particles. As such, they should not be interpreted in the same sense as the distance horizons reported in \citet{Kuznetsov2023} and \citet{Unger2024}, which consider the selection on energy independently. Particles originating at larger $D_\mathrm{src}$ experience larger deflections and are less likely to arrive from the same direction as Amaterasu. The arrival direction provides the strictest selection criterion, due to the relatively large uncertainties on Amaterasu's energy. Therefore, the effect of the arrival direction selection dominates and leads to relatively similar values of the distance at which the efficiency drops to below 10\% of its value at 1\,Mpc (used to define our $D_\mathrm{max}$) for cases 1) and 2), as shown in Fig.~\ref{fig:selection_efficiency}. However, an extension of the distribution to beyond $D_\mathrm{max} = 15$~Mpc cannot be ruled out based on this analysis. We focus on nearby sources in this work, but hope to explore larger volumes with similar methods in the future.

\begin{figure}[ht]
\includegraphics[width=\columnwidth]{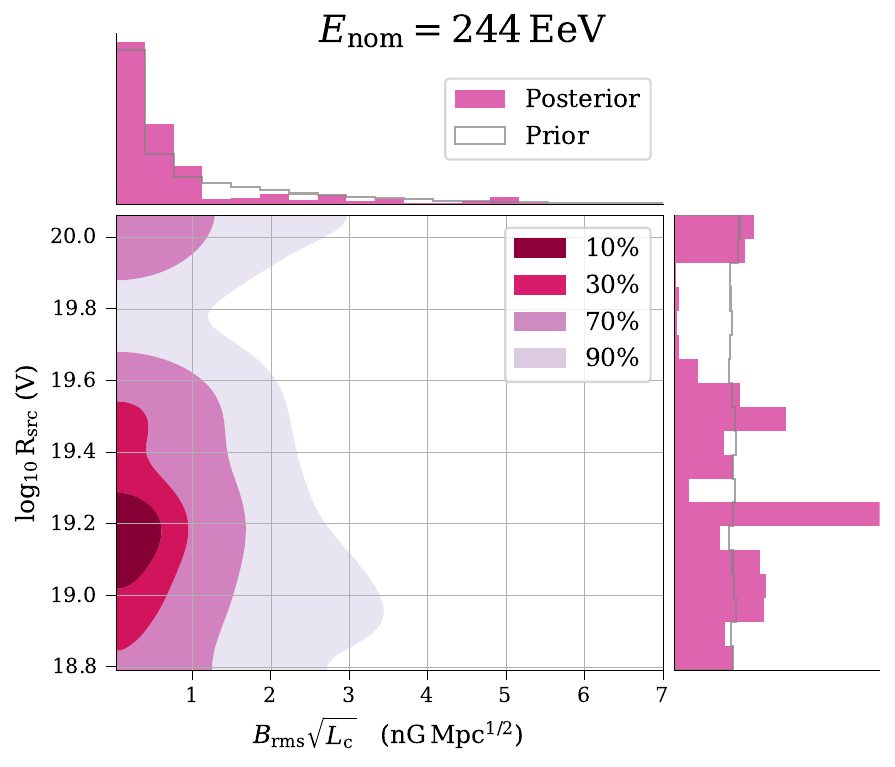}
\includegraphics[width=\columnwidth]{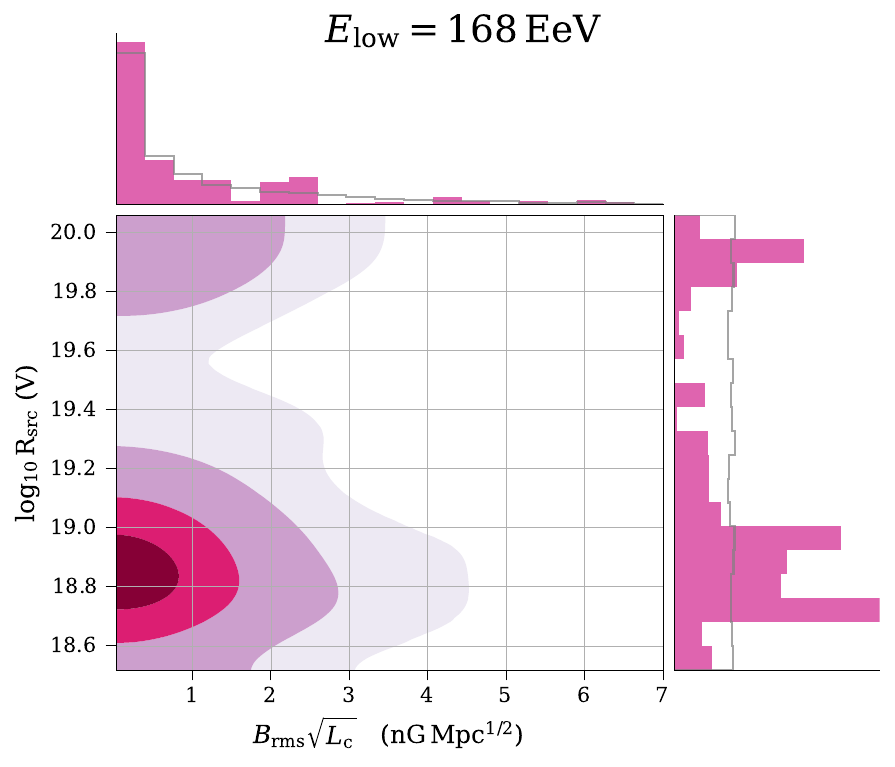}
\caption{The joint marginal posterior distribution for the source rigidity, $R_\mathrm{src}$, and the EGMF parameters, $B_\mathrm{rms} \sqrt{L_\mathrm{c}}$ for the nominal (upper) and low energy (lower) runs. The contours show the 10\%, 30\%, 70\%, and 90\% regions of highest posterior density. The upper and right panels of each plot show the marginal distributions of the prior and posterior for these parameters. \label{fig:rig_B}}
\end{figure}

We summarise the results for the remaining free parameters in Fig.~\ref{fig:rig_B}. To do so, we combine the EGMF parameters to reflect their impact on UHECR deflections and convert $E_\mathrm{src}$ into the source rigidity, ${R = E_\mathrm{src} /eZ}$, where $eZ$ is the particle charge. The preferred rigidity is $\mathrm{R\approx 10^{19.3}}$\,V for the nominal energy and $\mathrm{R\approx 10^{18.85}}$\,V for the low energy case. The results on $\mathrm{B_{rms}\sqrt{L_c}}$ show that we effectively marginalise over the prior, as expected.

\begin{figure}
    \centering
    \includegraphics[width=\columnwidth]{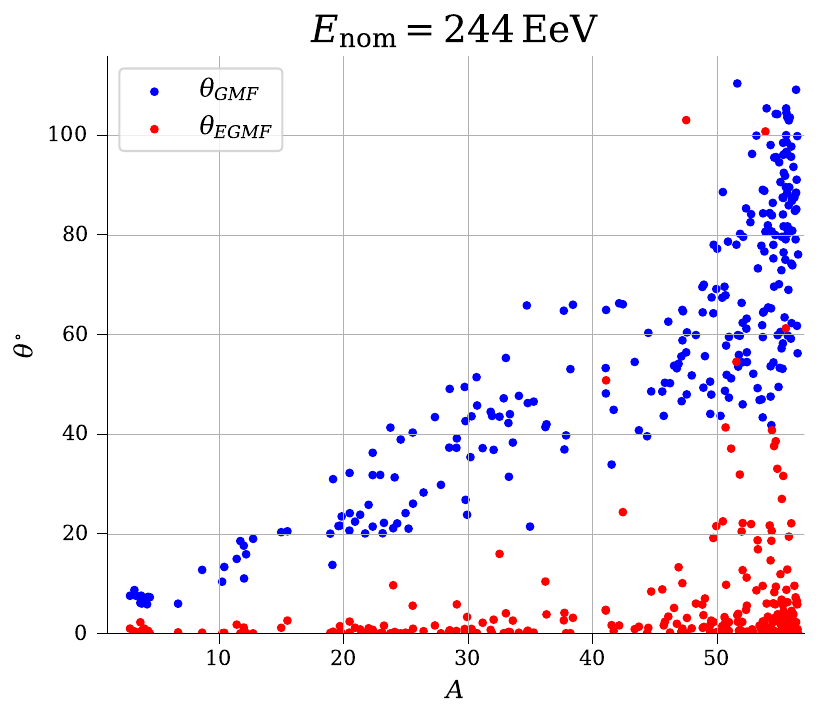}
    \includegraphics[width=\columnwidth]{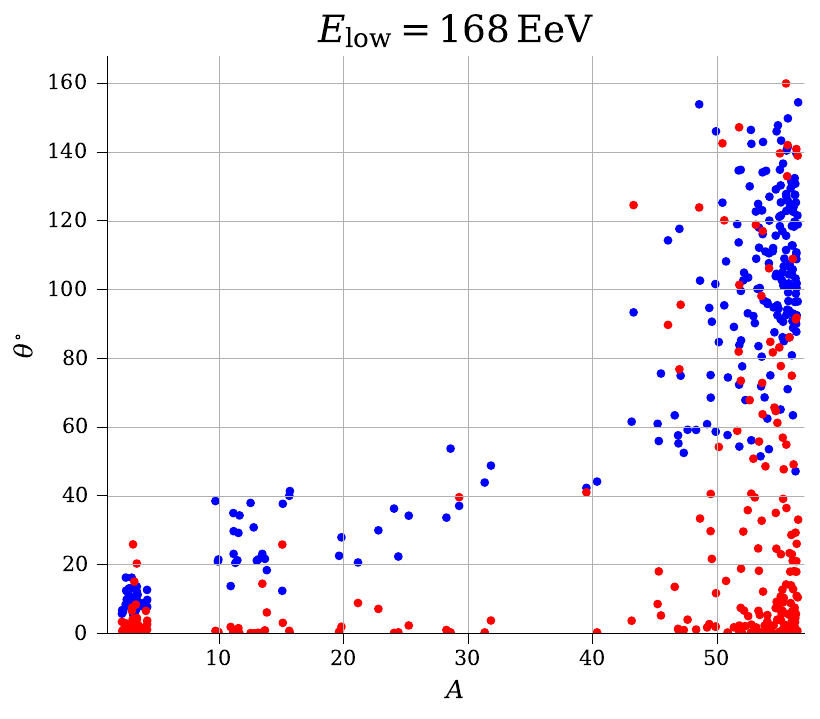}
    \caption{Deflection angles, $\theta$, of particles for all accepted parameter sets as a function of the arrival mass number, $A$. Deflections resulting from the GMF and EGMF are shown in green and red, respectively. The upper panel shows the case for $E_\mathrm{nom}$ and the lower panel shows the $E_\mathrm{low}$ case. For visibility, a random scatter sampled from a uniform distribution bounded by $\pm 0.5$ has been added to the $A$ values.}
    \label{fig:Bfield_angles}
\end{figure}

To demonstrate the relative impact of the GMF and EGMF assumptions on our results, we show the deflection angles as a function of arrival mass number for particles resulting from all accepted parameters sets in Fig.~\ref{fig:Bfield_angles}. We can see that the contribution of the GMF and EGMF is comparable for lighter and heavier arrival masses, but overall the impact of the GMF dominates. The EGMF deflections can be significantly larger, but only rarely and for heavier particles when considering the $E_\mathrm{low}$ case.   

Given that our assumption of a maximum source energy of up to 3\,ZeV is challenging from a theoretical perspective, we also test the impact of a lower energy cut on our results such that $E_\mathrm{src} \leq 500$\,EeV, as shown in Appendix~\ref{sec:lower_Emax}. For both case 1) and 2), the overall source direction distributions are slightly wider and there is a shift towards an even heavier arrival composition. As such, restricting the source energies to lower values still leads to the conclusion that Amaterasu's source location is consistent with regions outside of the Local Void.

\section{Discussion} \label{sec:sources}

Our combination of 3D \texttt{CRPropa 3} simulations with ABC allows us to include a more realistic physical model for UHECR propagation and directly constrain the parameters of this model and their correlations. 

When considering all possible arrival masses and $E_\mathrm{nom}$, as shown in the upper panel of Fig.~\ref{fig:Enom_skymap}, three main interesting astrophysical objects appear within the the 90\% contour of our volume, starburst galaxies NGC~6946 and M82, NGC~2403. Another galaxy contained in our contours is NGC~6789, an irregular dwarf galaxy. 

If we consider the composition-dependent contours, as shown in the lower panel of Fig.~\ref{fig:Enom_skymap}, we see that M82 and NGC~2403 are preferred in the case of a heavy arrival composition, with M82 being within the 70\% contour and NGC~2403 in the 90\% one, while NGC~6946 is consistent with both a heavy and a medium composition since it is found within the 30\% and 90\% contour of both volumes, respectively.

NGC~6946 has also been found close to the region of possible source positions in \citet{Unger2024}. However, it was disfavoured as a convincing source candidate as it contributes only 3\% to the total 1.4 GHz radio flux of SBGs within distances similar to those considered here, and the 1.4\,GHz radio flux is often used as a proxy of UHECR production \citep{PAO2018,Halim2024}.  

M82 is a powerful and nearby SBG at a distance of only 3.6 Mpc. It lies a few degrees from the TA hotspot and is commonly invoked as a UHECR source candidate \citep{TA2014, He2016}. M82 is also the brightest SBG within the source distances considered here, accounting for over 18\% of the total 1.4\,GHz radio flux \citep{PAO2018}. 

The lower energy case reveals several more objects consistent with our posterior distribution. In this case, shown in the upper panel of Fig.~\ref{fig:Elow_skymap}, our 10\% contour overlaps with multiple astrophysical objects that are within ${D_\mathrm{src} \leq 15}$~Mpc. However, most of these are galaxies appear relatively inactive and do not belong to a source class that satisfies the Hillas criterion \citep{Hillas1984}. An exception within the 10\% contour is NGC~7331, a starburst galaxy at 14.71 Mpc. The 70\% contour also contains NGC~6946, while the 90\% one contains NGC~891 and NGC~0660. However, NGC~7331, NGC~891 and NGC~0660 all have smaller 1.4\,GHz radio flux densities than NGC~6946 \citep{Lunardini2019}.

The mass-dependent posterior contours for the low-energy case reveal that NGC~7331, NGC~891, NGC~0660 and NGC~6946 are consistent with the case of a heavier arrival composition. NGC~6946 is also favoured for the case of a medium arrival composition, since it is found within the 30\% highest density region. Additionally for this case, NGC~1569 and IC~342 are found within the 90\% contours. Maffei~2 lies at the edge of the 90\% contours for both medium and heavy arrival compositions.

Our results offer alternatives to the origins of Amaterasu outside of the Local Void when considering key modelling uncertainties, the possibility of a heavier arrival mass, and/or the lower end of the systematic energy uncertainty. While there are currently no strong constraints on the arrival mass composition, it seems that the UHECR composition evolves with energy and we can expect to see heavier elements at the highest energies \citep{Auger:2025ks}. 

There are several possible phenomena that can drive UHECR acceleration in SBGs, such as superwinds caused by the additive action of stellar winds and supernovae \citep{Anchordoqui2019}. Furthermore, SBGs are rich in young, massive stars which leads to an increased rate of long-duration gamma-ray bursts \citep{Marafico:2024ps}. Another explanation for the association between UHECRs and SBGs that also includes the action of AGN is offered by the echo model \citep{Bell2022}.

Alternatively, Amaterasu could have been produced by a past transient event. The average propagation time of our accepted events for the nominal and the low energy case throughout Galactic and extra-Galactic magnetic fields is ${\approx 10}$~Myrs and ${\approx 50}$~Myrs, respectively. In addition to long-duration gamma-ray bursts, transient events that could be responsible for UHECR acceleration are tidal disruption events, or binary neutron star mergers \citep{Zhang2017,Biehl2018,Batista2019,Farrar2024,Bartos:2025sg}.

The main differences between our work and the results reported in \citet{Unger2024} are: 1) the ABC approach including 3D propagation modelling and a joint selection on the arrival energy and direction of Amaterasu, 2) the ability to condition on the arrival particle type, 3) the inclusion of an effective EGMF up to values of $B_\mathrm{rms} \sqrt{L_c} < 10^{-8}$~G $\mathrm{Mpc}^{1/2}$, and 4) the use of the UF23 base model for the coherent GMF as opposed to the full suite of 8 UF23 models. These important distinctions mean that it is challenging to directly compare the results as the analyses involve different assumptions and the reported constraints on the source location have a fundamentally different meaning. 

With this in mind, for the case of $E_\mathrm{low}$ and the assumption of a heavy arrival composition where our assumptions are closest to those of \citet{Unger2024}, we find that our 70\% region of highest posterior density for the source direction is broadly consistent with the various contours reported in their Fig.~5. The 90\% region that we also report covers a larger region, which can be understood both in the different definition of uncertainty and the inclusion of the EGMF in our modelling. As shown in Fig.~\ref{fig:Bfield_angles}, the impact of the EGMF can be comparable to the GMF for the particles resulting from our accepted parameter sets and can therefore have an non-negligible impact on the tails of the source direction distributions. While we do not yet include the systematic uncertainty of the GMF modelling in our work, we expect this to lead to further broadening of the source direction distribution, as shown in \citet{Unger2024}. 

In this work, we only consider distances out to 15\,Mpc. As shown in Fig.~\ref{fig:selection_efficiency}, the acceptance criterion for the particle arrival direction restricts the efficiency of our algorithm, making an exploration of larger distances computationally challenging. However, we focus here on demonstrating our approach by exploring possible astrophysical source candidates in the nearby Universe. We expect that increasing $D_\mathrm{max}$ to give a more comprehensive picture of the source distance posterior will only lead to further possible source candidates, also from the increased impact of the EGMF further opening up the posterior volume.  

M82 stands out as an interesting source candidate that has not been highlighted in previous works given the distance and direction constraints for the $E_\mathrm{nom}$ case and a heavier arrival composition (see Fig.~\ref{fig:Enom_skymap}). TA reports that a systematic uncertainty of -10\% on the energy scale should be considered for the unknown primary composition \citep{TA2023}. While this falls within the $-3\sigma_E$ statistical uncertainties considered for the arrival energy acceptance criterion in this work, we note that M82 does not fall within the 90\% contours for the $E_\mathrm{low}$ case, regardless of our assumptions regarding the arrival particle type. This result remains unchanged when considering a lower maximum source energy of 500~EeV, as detailed in Appendix~\ref{sec:lower_Emax}.

\section{Conclusions} \label{sec:conclusion}

We have developed a new method to map out the volume of space consistent with the origins of individual particles at the highest energies. Our ABC-based approach allows us to include a full 3D model for UHECR propagation with Galactic and extra-Galactic magnetic fields, as well as key sources of uncertainty. The results are presented in the form of posterior distributions over the 6 free model parameters: $l_\mathrm{src}$, $b_\mathrm{src}$, $D_\mathrm{src}$, $E_\mathrm{src}$, $B_\mathrm{rms}$ and $L_c$. The constraints on $l_\mathrm{src}$, $b_\mathrm{src}$ and $D_\mathrm{src}$ provide an interpretable probability density map for the most likely origin of the analysed particle. This result is further supported by physical constraints on $E_\mathrm{src}$ and correlated uncertainties that are marginalised over through the incorporation of a prior on the EGMF properties, $B_\mathrm{rms}$ and $L_c$.

In this work, we demonstrate our method through application to the interesting case of the Amaterasu particle detected by TA. Assuming that the particle is an iron nucleus at the source, we find that its origins are consistent with regions of space beyond the Local Void. The largest regions that are most shifted from the arrival direction of Amaterasu are found for the case of medium to heavy arrival particle masses and the $E_\mathrm{low}$ case, as expected due to the increased impact of magnetic fields. Several astrophysical source candidates lie within the proposed volumes, including SBGs and quiescent galaxies, which could also host past transient events capable of energetic particle acceleration. 

Our approach goes beyond previous studies by addressing the joint selection on arrival direction and energy and the correlated constraints this implies for the possible source distance. Furthermore, we include the impact of the EGMF and the ability to condition on the arrival particle type. As such, we find several new source candidates, notably M82. However, the current implementation of our ABC algorithm is computationally inefficient, meaning that we are limited to the assumption that Amaterasu is an iron nucleus at its source, we only explore possible source distances out to 15\,Mpc and our prior for the source direction cannot be uniform over the whole sky. We plan to address these challenges in future work within the framework of simulation-based inference, to allow for a more comprehensive studies and extension to a larger number of UHECRs at the highest energies. More detailed studies of the impact of the source direction prior and the GMF modelling assumptions are also planned in this context.

Our results highlight the importance of resolving the UHECR energy and arrival mass or charge at the highest energies. We see a significant impact on the possible volumes consistent with the Amaterasu particle for the different cases considered here. For example, a more constrained measurement of the nominal energy reported for Amaterasu or evidence for a lighter arrival composition at these energies would leave little room for origins outside of the Local Void. Planned upgrades to existing observatories such as TAx4 \citep{Fujisue2023} and AugerPrime \citep{Castellina2019} will soon augment our knowledge, alongside improved data analyses methods leveraging neural networks to give new constraints on the UHECR spectrum and its composition \citep{Auger:2025ks}. We also expect further advances from next-generation facilities such as POEMMA and GRAND \citep{Coleman2023}. 

\section*{Acknowledgments}
We thank M. Unger, T. K. Bister, and A. van Vliet for their valuable input during our discussions. We also thank A. Fedynitch and K. Watanabe for the UF23 model lenses. We acknowledge the Max Planck Computing and Data Facility for the use of the Raven and Viper HPC systems. N. Bourriche acknowledges the financial support from the Excellence Cluster ORIGINS, which is funded by the Deutsche Forschungsgemeinschaft (DFG, German Research Foundation) under Germany’s Excellence Strategy - EXC-2094-390783311.

\software{\texttt{astropy} \citep{2013A&A...558A..33A,2018AJ....156..123A},
  \texttt{numpy} \citep{numpy}, \texttt{matplotlib}
  \citep{Hunter:2007}, \texttt{scipy} \citep{2020SciPy-NMeth},
  \texttt{h5py} \citep{collette_python_hdf5_2014,h5py_7560547},
  \texttt{seaborn} \citep{Waskom2021}, \texttt{cartopy} \citep{Cartopy}, 
  \texttt{CRPropa 3} \citep{Batista2022}}

\bibliography{main}{}
\bibliographystyle{aasjournal}

\appendix

\section{Lower maximum energy at the source}
\label{sec:lower_Emax}

\begin{figure*}[ht]
\centering
\includegraphics[width=0.9\textwidth]{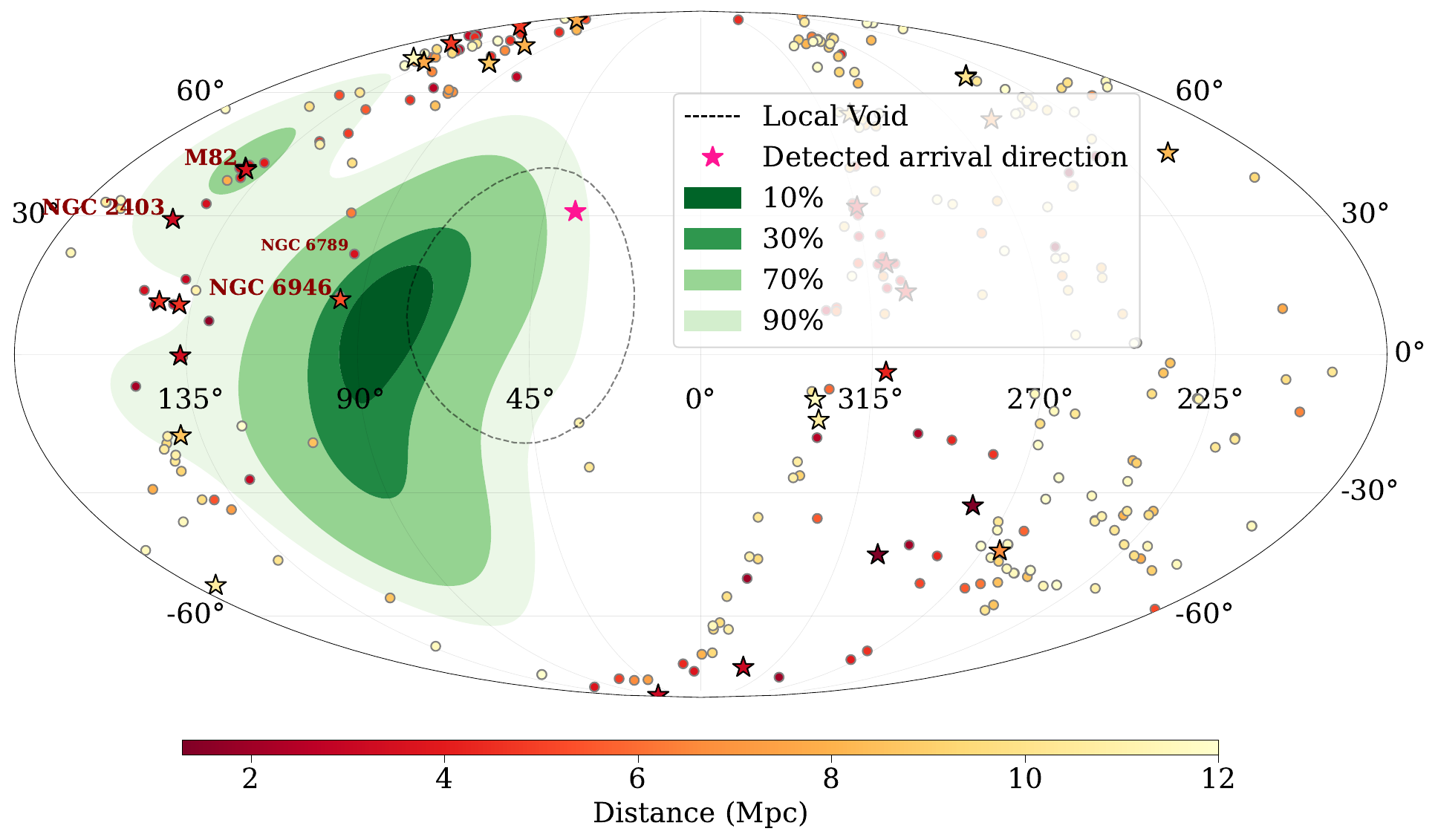}
\includegraphics[width=0.9\textwidth]{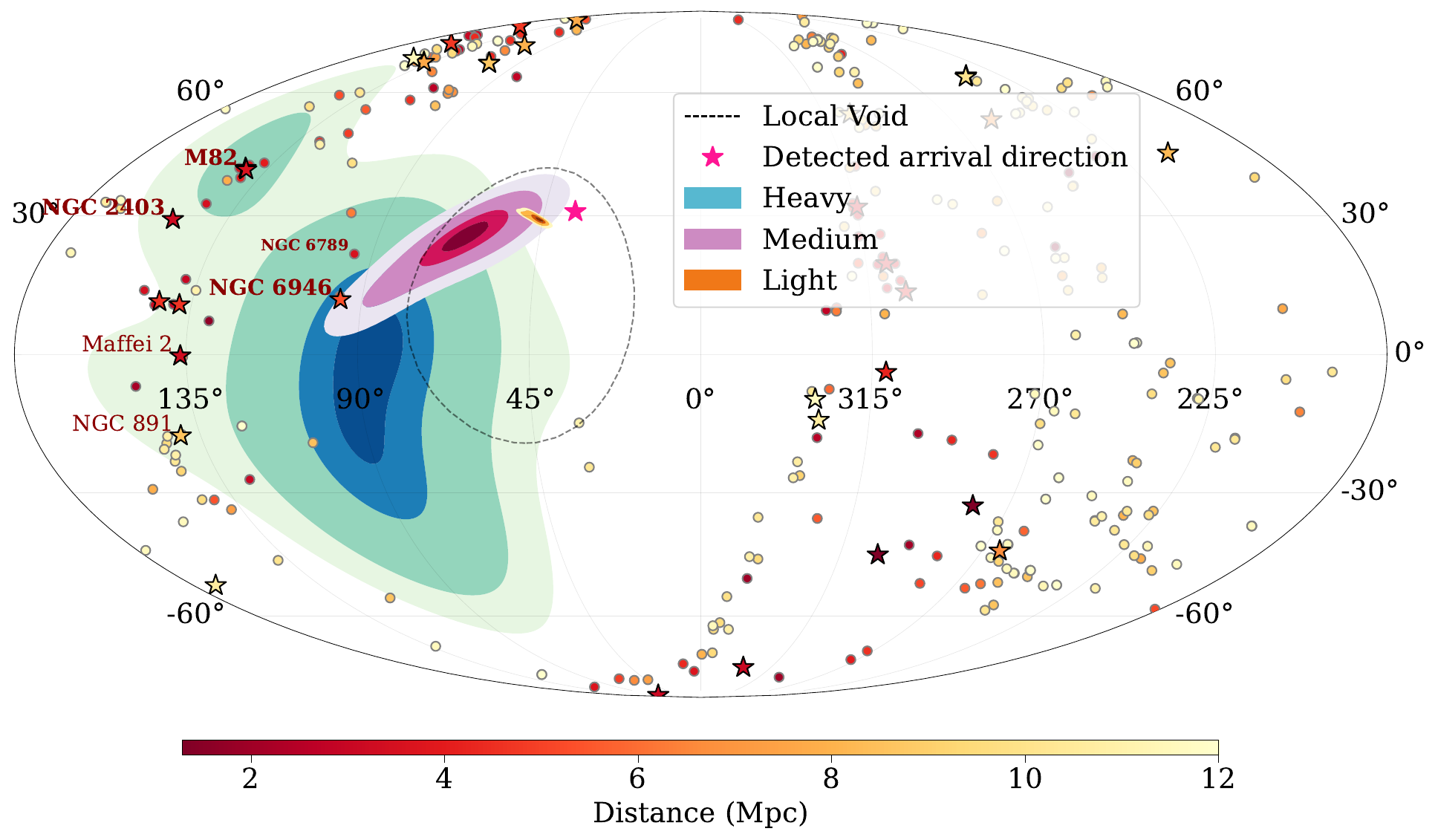}
\caption{Sky maps for the $E_\mathrm{nom}$ case with an extra energy cut such that $E_\mathrm{src} \leq 500$\,EeV. The possible source directions are shown in Galactic coordinates. The measured arrival direction of Amaterasu, and the outline of the Local Void are shown for reference. The markers show galaxies within the accepted $D_\mathrm{src}$ range, with stars indicating SBGs and AGN and circles showing quiescent galaxies. In the upper panel, all accepted parameter sets are considered and the green contours outline the labelled regions of highest posterior density. The lower panel shows the arrival mass-dependent posterior distribution, conditional on the particle arriving with a mass number, $A$, in one of three groups, as shown in the legend. The contour levels are as in the upper panel.} \label{fig:Enom_skymap_Ecut}
\end{figure*}

\begin{figure*}[ht]
\centering
\includegraphics[width=0.9\textwidth]{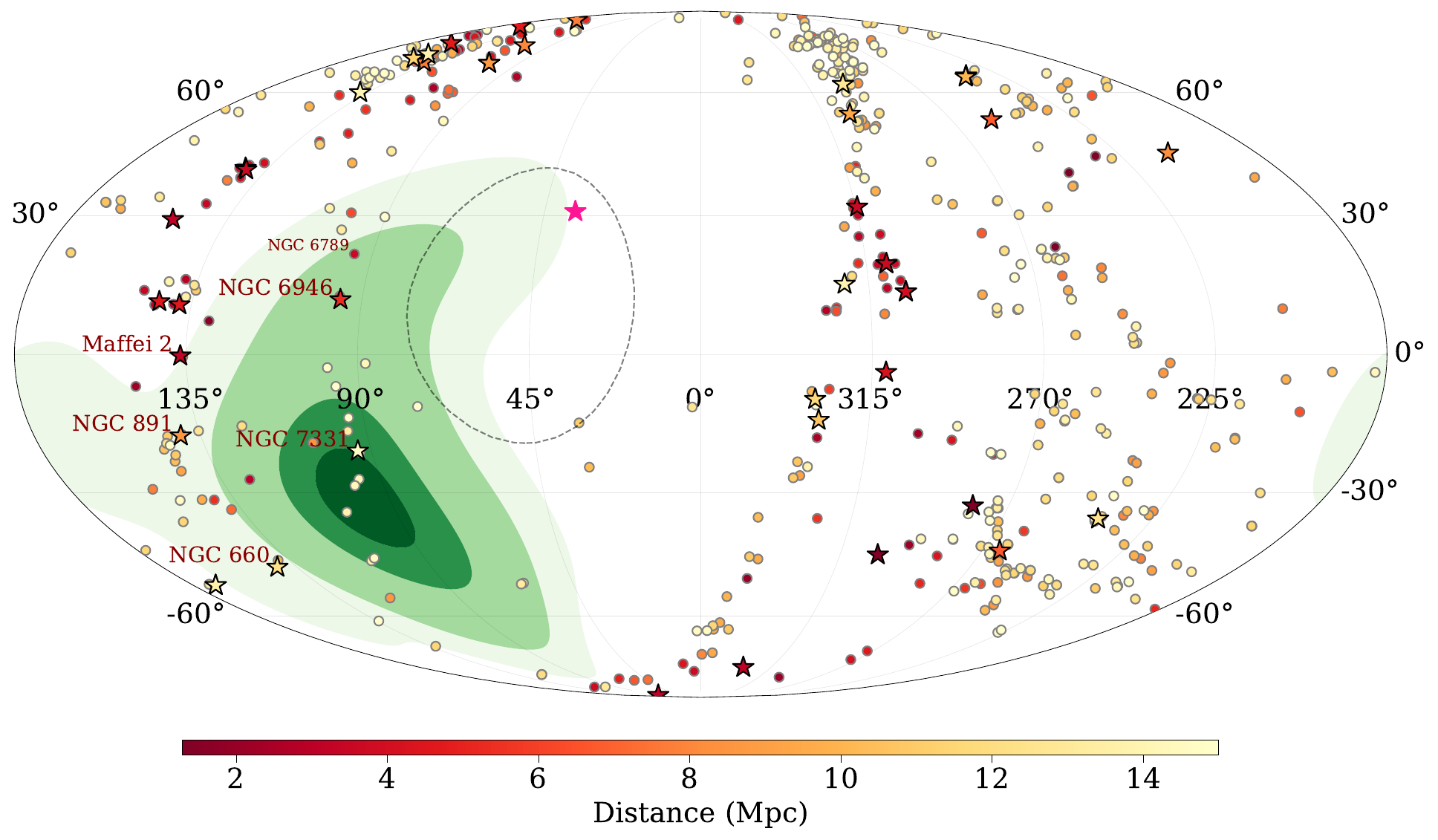}
\includegraphics[width=0.9\textwidth]{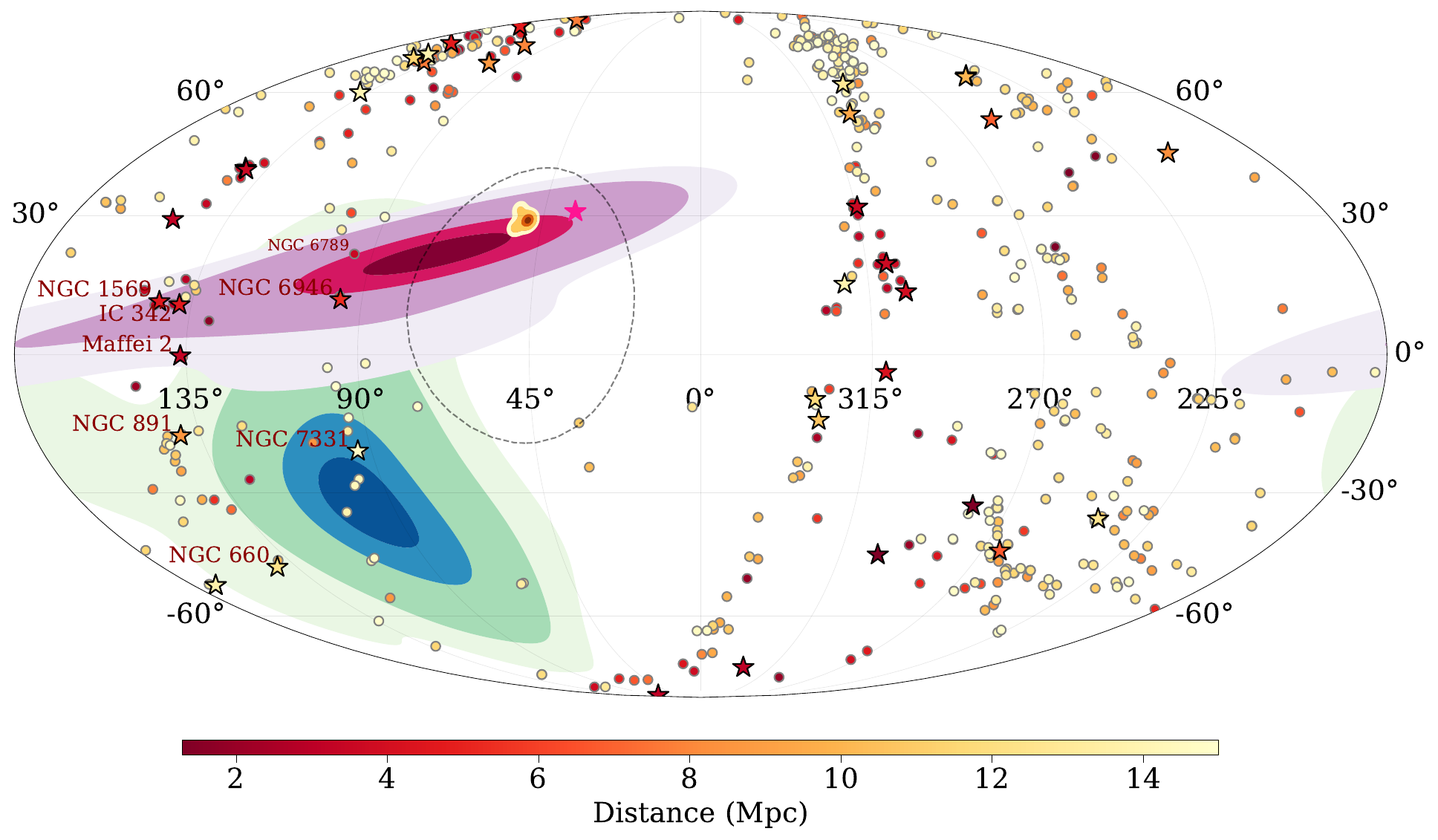}
\caption{Sky maps for the $E_\mathrm{low}$ case with an extra energy cut such that $E_\mathrm{src} \leq 500$\,EeV. The figure layout is as in Fig.~\ref{fig:Enom_skymap_Ecut}.} \label{fig:Elow_skymap_Ecut}
\end{figure*}

\begin{figure}[ht]
\centering
\includegraphics[width=0.47\textwidth]{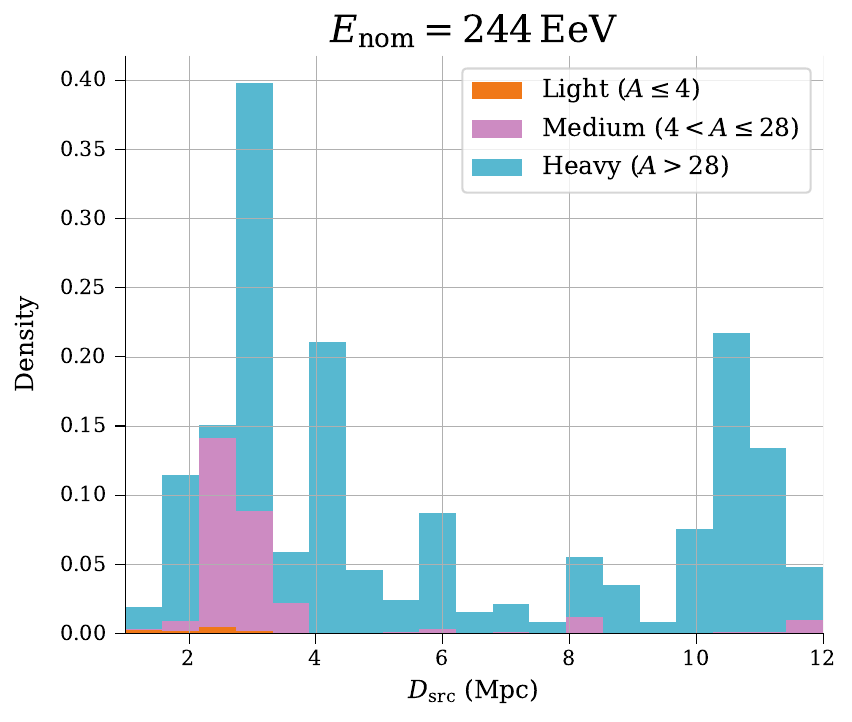}
\includegraphics[width=0.47\textwidth]{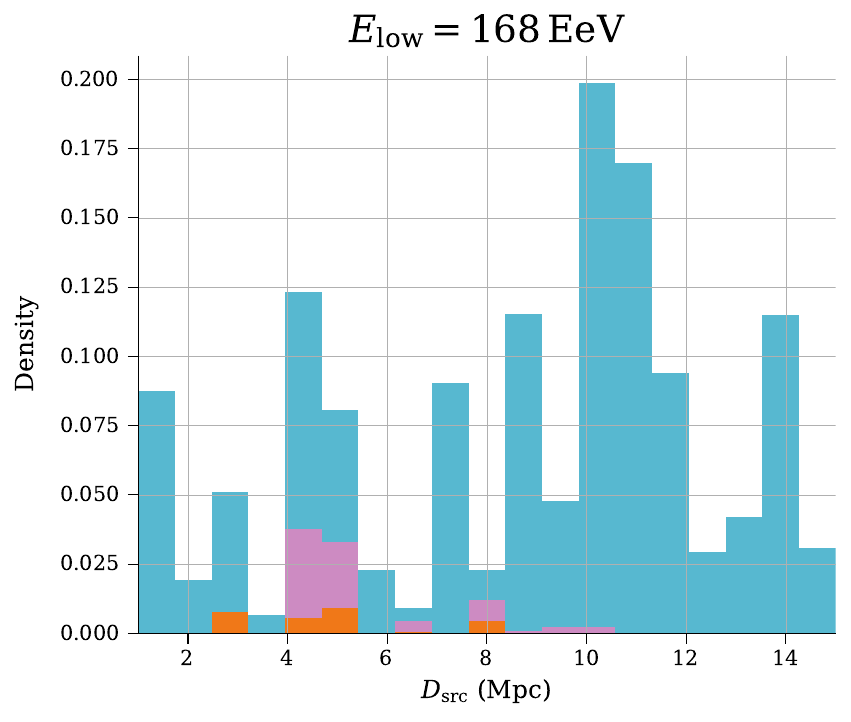}
\caption{The marginal posterior distribution of $D_\mathrm{src}$ is shown for for the case with an extra energy cut such that $E_\mathrm{src} \leq 500$\,EeV. The nominal and low energy cases are shown in the left and right panels, respectively. The results are shown as a stacked histogram to highlight the relative contributions of $D_\mathrm{src}$ values that lead to accepted particles in different arrival mass groups. Orange, pink, and blue bars indicate arrival masses that are light, medium and heavy, respectively. The height of the bar gives the total density summed over all mass numbers and the histogram is normalised. \label{fig:Ecut_D_hist}}
\end{figure}

\begin{figure}[ht]
\centering 
\includegraphics[width=0.47\textwidth]{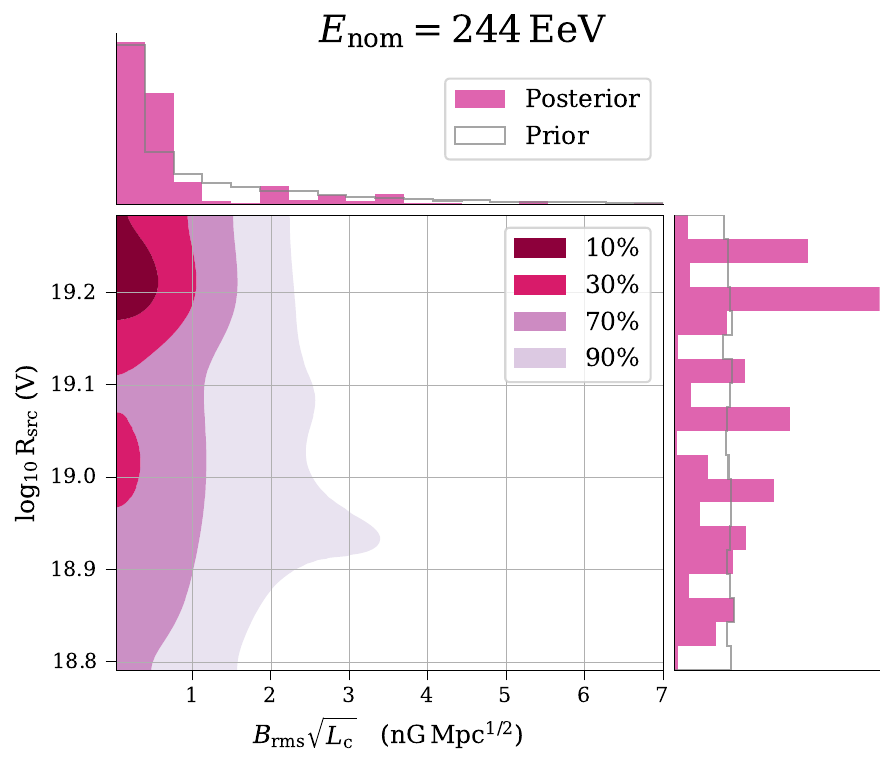}
\includegraphics[width=0.47\textwidth]{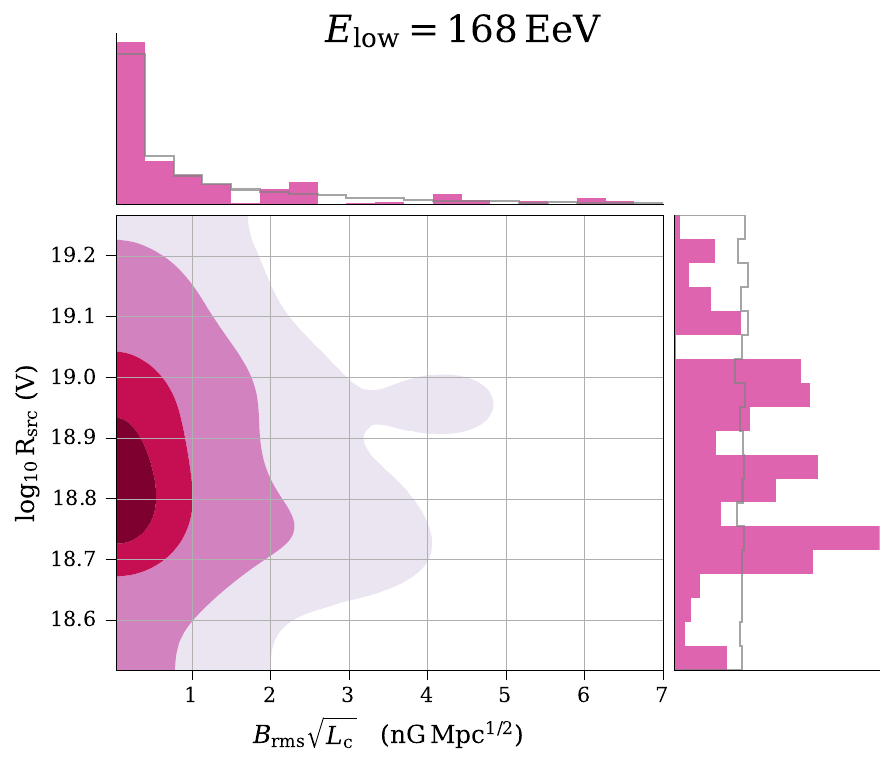}
\caption{The joint marginal posterior distribution for the source rigidity, $R_\mathrm{src}$, and the EGMF parameters, $B_\mathrm{rms} \sqrt{L_\mathrm{c}}$ for the case with an extra energy cut such that $E_\mathrm{src} \leq 500$\,EeV. The left and right panels show the  nominal and low energy cases, respectively. We note the different y-axis limits compared to Fig.~\ref{fig:rig_B} due to the use of a different energy range. \label{fig:Ecut_R_vs_B}}
\end{figure}

We apply a cut for $E_\mathrm{src} \leq 500$\,EeV to explore the effects of a lower source rigidity assumption on our results. In Figs.~\ref{fig:Enom_skymap_Ecut} and \ref{fig:Elow_skymap_Ecut}, we show the sky maps for $E_\mathrm{nom}$ and $E_\mathrm{low}$ for all arrival mass groups combined and also split into light, medium and heavy arrival compositions, as described in Section~\ref{sec:results} and Figs.~\ref{fig:Enom_skymap} and \ref{fig:Elow_skymap}. The majority (around $\sim 65$\,\%) of particles resulting from accepted parameter sets for both $E_{\text{nom}}$ and $E_{\text{low}}$ have $E_\mathrm{src} \leq 500$\,EeV. In both cases, the arrival composition is heavier than compared to the case of $E_\mathrm{src} \leq 3000$\,EeV, with over 80\% of the events belonging to the third mass group. This larger heavy contribution means that the contours for the light and medium compositions slightly shrink while the contours for the heavy composition remain almost unchanged. Fig.~\ref{fig:Ecut_D_hist} shows this more clearly; the majority of the accepted trials are heavy nuclei, especially for the $E_{\text{low}}$ case. This shows that higher maximum energies are required to allow a significant light contribution give the assumption of iron at the source along with the priors described in Section~\ref{sec:simset}. We also show the joint posterior distribution of $R_\mathrm{src}$ and $B_\mathrm{rms}\sqrt{L_c}$ in Fig.~\ref{fig:Ecut_R_vs_B}. Here, there is a shift to lower rigidities due to the imposed cut, and the lack of constraints on the EGMF remains unchanged.

\begin{table}
\begin{tabular}{lllll}
\toprule
& \textbf{Total} & \textbf{Light} & \textbf{Medium} & \textbf{Heavy} \\
\midrule
$E_\mathrm{nom} (E_\mathrm{src} \leq 500$\,EeV) & 176 & 3.4\,\% & 11.9\,\% & 84.7\,\% \\
$E_\mathrm{low} (E_\mathrm{src} \leq 500$\,EeV) & 160 & 3.7\,\% & 9.4\,\% & 86.9\,\% \\
\bottomrule
\end{tabular}
\caption{Number of accepted parameter sets resulting from running ABC, as shown in Table~\ref{tab:acc_param_sets}. This time the results are shown including a cut on the maximum energy of $E_\mathrm{src} \leq 500$\,EeV. \label{tab:acc_param_sets_Ecut}}
\end{table}

\end{document}